\documentclass[12pt]{article}
\usepackage{amssymb}
\renewcommand{\theequation}{\arabic{section}.\arabic{equation}}

\font\oneeight=cmr10 at 18pt
\font\onefour=cmr10 at 14pt

\newcommand{\vTm}{\vphantom{\mbox{\oneeight I}}}
\newcommand{\vTs}{\vphantom{\mbox{\onefour I}}}

\begin{document}

%************************** Text Begins here ******************************

%  Greek letters

\def\a{\alpha}
\def\b{\beta}
\def\d{\delta}
\def\e{\epsilon}
\def\g{\gamma}
\def\h{\mathfrak{h}}
\def\k{\kappa}
\def\l{\lambda}
\def\o{\omega}
\def\t{\theta}
\def\s{\sigma}
\def\x{\xi}
  \def\A{{\cal{A}}}
  \def\B{{\cal{B}}}
  \def\C{{\cal{C}}}
  \def\D{{\cal{D}}}
\def\G{\Gamma}
\def\O{\Omega}
\def\L{\Lambda}
\def\f{E_{\tau,\eta}(sl_2)}
\def\E{E_{\tau,\eta}(sl_n)}
\def\r{E_{\tau,\eta}(sl_{n-1})}
\def\Zb{\mathbb{Z}}
\def\Cb{\mathbb{C}}

\def\R{\overline{R}}
% Shorthands for \begin{equation} and the like

\def\beq{\begin{equation}}
\def\eeq{\end{equation}}
\def\bea{\begin{eqnarray}}
\def\eea{\end{eqnarray}}
\def\ba{\begin{array}}
\def\ea{\end{array}}
\def\no{\nonumber}
\def\le{\langle}
\def\re{\rangle}
\def\lt{\left}
\def\rt{\right}

\newtheorem{Theorem}{Theorem}
\newtheorem{Definition}{Definition}
\newtheorem{Proposition}{Proposition}
\newtheorem{Lemma}{Lemma}
\newtheorem{Corollary}{Corollary}
\newcommand{\proof}[1]{{\bf Proof. }
         #1\begin{flushright}$\Box$\end{flushright}}

\baselineskip=20pt

%%%%%%%%%%%%%%%%%%%%%%%%%%%%%%%%%%%%%%%%%%%%%%%%%%%%%%%%%%%%
%                                                          %
%  Title page                                              %
%                                                          %
%%%%%%%%%%%%%%%%%%%%%%%%%%%%%%%%%%%%%%%%%%%%%%%%%%%%%%%%%%%%
\newfont{\elevenmib}{cmmib10 scaled\magstep1}
\newcommand{\preprint}{
    \begin{flushleft}
      \elevenmib Yukawa\, Institute\, Kyoto\\
    \end{flushleft}\vspace{-1.3cm}
    \begin{flushright}\normalsize  \sf
      YITP-03-06\\
      {\tt hep-th/0303077} \\ March  2003
    \end{flushright}}
\newcommand{\Title}[1]{{\baselineskip=26pt
    \begin{center} \Large \bf #1 \\ \ \\ \end{center}}}
\newcommand{\Author}{\begin{center}
    \large \bf B.\,Y.~Hou${}^a$, ~R.~Sasaki${}^b$~ and~W.-L. Yang${}^{a,b}$
\end{center}}
\newcommand{\Address}{\begin{center}
      $^a$ Institute of Modern Physics, Northwest University\\
      Xian 710069, P.R. China\\
      ~~\\
      ${}^b$ Yukawa Institute for Theoretical Physics,\\
      Kyoto University, Kyoto 606-8502, Japan
    \end{center}}
\newcommand{\Accepted}[1]{\begin{center}
    {\large \sf #1}\\ \vspace{1mm}{\small \sf Accepted for Publication}
    \end{center}}

\preprint
\thispagestyle{empty}
\bigskip\bigskip\bigskip

\Title{Algebraic Bethe ansatz  for the elliptic quantum group
  $E_{\tau,\eta}(sl_n)$ and its applications
} \Author

\Address
\vspace{1cm}

\begin{abstract}
We study the tensor product of the {\it higher spin
representations} (see the definition in Sect. 2.2) of the elliptic
quantum group $\E$.  The  transfer matrices associated  with the
$\E$-module are exactly diagonalized  by the nested Bethe ansatz
method. Some special cases of the construction give the exact solution
for the $\Zb_n$ Belavin model and for the elliptic $A_{n-1}$
Ruijsenaars-Schneider model.

\vspace{1truecm} \noindent {\it PACS:} 03.65.Fd; 05.30.-d

\end{abstract}
\newpage
%%%%%%%%%%%%%%%%%%%%%%%%%%%%%%%%%%%%%%%%%%%%%%%%%%%%%%%%%%%%%%%
%                                                             %
%  1. Introduction                                            %
%                                                             %
%%%%%%%%%%%%%%%%%%%%%%%%%%%%%%%%%%%%%%%%%%%%%%%%%%%%%%%%%%%%%%%
\section{Introduction}
\label{intro}
\setcounter{equation}{0}

Bethe ansatz method  has  proved to be the most powerful and
(probably) unified method to construct the common eigenvectors of
commuting families of operators (usually called {\it  transfer
matrices}) in two-dimensional integrable models \cite{Fad79,
Tha82, Kor93}. Faddeev et al \cite{Fad79} reformulated the Bethe
ansatz method  in a representation theory form:  {\it transfer
matrices} are associated with  representations of certain algebras
with quadratic relations (now called quantum groups). The
eigenvectors are constructed by  acting certain algebra elements
on the `` highest weight vectors". This type of  Bethe ansatz
is known as  the algebraic Bethe ansatz.

Whereas this construction has been very successful in rational and
trigonometric integrable models \cite{Bab82, Sch83}, its extension
to elliptic models had been problematic due to the fact that for
the underlying algebras --$\Zb_n$ Sklyanin algebras \cite{Skl83,
Hou891} the highest weight representations were not properly
defined. Therefore, the algebraic Bethe ansatz method had not be
applied to the elliptic integrable models directly.

Recently, a definition of elliptic quantum groups
$E_{\tau,\eta}(\mathfrak{g})$ associated with any simple classical
Lie algebra $\mathfrak{g}$ was given \cite{Fel94}. The highest
weight  representations of the elliptic quantum groups (cf.
$\Zb_n$ Sklyanin algebras) are now well-defined \cite{Fel94,
Fel961}. This enabled Felder et al \cite{Fel96} to apply
successfully  the algebraic Bethe ansatz method for constructing
the eigenvectors of the {\it transfer matrices\/} of the elliptic
integrable models associated with   modules over
$E_{\tau,\eta}(sl_2)$, and Billey \cite{Bil98} to apply the
algebraic nested Bethe ansatz method for constructing the
eigenvalues of the {\it transfer matrices\/} associated with a
special module over $\E$ (see section 5).

In this paper, we will extend the above construction of Bethe
ansatz method  further to the elliptic quantum group $\E$ on a
generic {\it higer spin\/} module $W=V_{\L^{(l_1)}}(z_1)\otimes
V_{\L^{(l_2)}}(z_2)\otimes \cdots\otimes V_{\L^{(l_m)}}(z_m)$.
After briefly reviewing the definition of the elliptic quantum
group $\E$, we study  one parameter series of highest weight
representations ({\it higher spin representations}) of $\E$, and
introduce the associated operator algebra and the {\it transfer
matrices} corresponding to the $\E$-module in section 2. In
section 3, we describe the algebraic Bethe ansatz for $\E$.
Finally, we give som e applications of our construction to the
$\Zb_n$ Belavin models in section 4, and the elliptic
Ruijsenaars-Schneider model associated to $A_{n-1}$ root system in
section 5.

%%%%%%%%%%%%%%%%%%%%%%%%%%%%%%%%%%%%%%%%%%%%%%%%%%%%%%%%%%%%%%%
%                                                             %
%  2.The elliptic quantum group associated to                 %
%         $A_{n-1}$ type root system                          %
%                                                             %
%%%%%%%%%%%%%%%%%%%%%%%%%%%%%%%%%%%%%%%%%%%%%%%%%%%%%%%%%%%%%%%
\section{The elliptic quantum group and its modules associated to $A_{n-1}$}
\label{EQG} \setcounter{equation}{0}
\subsection{The elliptic quantum group associated to $A_{n-1}$}
We first review the definition of the elliptic quantum group $\E$
associated to $A_{n-1}$ \cite{Fel94}. Let
$\lt\{\e_{i}~|~i=1,2,\cdots,n\rt\}$ be the orthonormal basis of
the vector space $\Cb^n$ such that $\langle\e_i,~\e_j
\rangle=\d_{ij}$. The $A_{n-1}$ simple roots are
$\lt\{\a_{i}=\e_i-\e_{i+1}~|~i=1,\cdots,n-1\rt\}$ and the
fundamental weights $\lt\{\L_i~|~i=1,\cdots,n-1\rt\}$ satisfying
$\langle\L_i,~\a_j\rangle=\d_{ij}$ are given by
\bea
\L_i=\sum_{k=1}^{i}\e_k-\frac{i}{n}\sum_{k=1}^{n}\e_k. \no
\eea
Set
\bea
\hat{i}=\e_i-\overline{\e},~~\overline{\e}=
\frac{1}{n}\sum_{k=1}^{n}\e_k,~~i=1,\cdots,n,~~{\rm
then}~~\sum_{i=1}^n\hat{i}=0. \label{Vectors}
\eea
For each
dominant weight $\L=\sum_{i=1}^{n-1}a_i\L_{i}~,~~a_{i}\in \Zb^+$,
there exists an irreducible highest weight finite-dimensional
representation $V_{\L}$ of $A_{n-1}$ with the highest vector $
|\L\rangle$. For example the fundamental vector representation is
$V_{\L_1}$. In this paper, we only consider the symmetric
tensor-product representation of
$\stackrel{l}{\overbrace{V_{\L_1}\otimes V_{\L_1}\cdots \otimes
V_{\L_1}}}$ (or, the {\it higher spin-$l$ representation} of
$A_{n-1}$), namely, the one parameter series of highest weight
representations $V_{\L^{(l)}}$, with
\bea
\L^{(l)}=l\L_1,~~l\in\Zb~~{\rm and~~} l>0\label{Weight}.
\eea
This corresponds to the Young diagram
$\stackrel{l}{\overbrace{\Box\!\Box\!\Box\cdots\Box}}$.

Let $\h$ be the Cartan subalgebra of $A_{n-1}$ and $\h^{*}$ be its
dual. A finite dimensional diagonalizable  $\h$-module is a
complex finite dimensional vector space $W$ with a weight
decomposition $W=\oplus_{\mu\in \h^*}W[\mu]$, so that $\h$ acts on
$W[\mu]$ by $x\,v=\mu(x)\,v$, $(x\in \h,~v\in W[\mu])$. For
example, the fundamental vector representation $V_{\L_1}=\Cb^n$,
the non-zero weight spaces are $W[\hat{i}]=\Cb \e_i,~i=1,\cdots,n$.

Let us fix $\tau$ such that $Im(\tau)>0$ and a generic complex
number $\eta$. For convenience, we introduce another parameter
$w=n\eta$ related to $\eta$.  Let us introduce  the following
elliptic functions
\bea &&
\s(u)=\t\lt[\begin{array}{c}\frac{1}{2}\\[2pt]
\frac{1}{2}
\end{array}\rt](u,\tau),~~~~
\t^{(j)}(u)=\t\lt[\begin{array}{c}\frac{1}{2}-\frac{j}{n}\\\frac{1}{2}
\end{array}\rt](u,n\tau),
\label{Function}\\
&&\t\lt[
\begin{array}{c}
a\\b
\end{array}\rt](u,\tau)=\sum_{m=-\infty}^{\infty}
\exp{\lt\{\sqrt{-1}\pi\lt[(m+a)^2\tau+2(m+a)(u+b)\rt]\rt\}}.
\eea

For a generic complex weight $\x\in \sum_{i=1}^{n-1}\Cb \L_i$,  we
introduce a  weight lattice with a shift by $\x$, $\mathbb{P}_{\x}$ as
follows
\bea
\mathbb{P}_{\x}=\x+\sum_{i=1}^{n}m_i\hat{i},~~~m_i\in\Zb.
\label{Def}
\eea
For $\l\in \mathbb{P}_{\x}$, define
\bea
\l_i=\langle\l,\e_i\rangle,
~~\l_{ij}=\l_i-\l_j=\langle\l,\e_i-\e_j\rangle,~~i,j=1,\cdots,n.
\label{Def1}
\eea
In this paper, we restrict  $\l\in
\mathbb{P}_{\x}$ so that the inverse of the matrix $S(z,\l)$ in
(\ref{twisting}) does exist. Let $R(z,\l)\in
End(\Cb^n\otimes\Cb^n)$ be the R-matrix given by
\bea
&&R(z,\l)=\sum_{i=1}^{n}R^{ii}_{ii}(z,\l)E_{ii}\otimes E_{ii}
+\sum_{i\ne j}\lt\{R^{ij}_{ij}(z,\l)E_{ii}\otimes E_{jj}+
R^{ji}_{ij}(z,\l)E_{ji}\otimes E_{ij}\rt\},\no\\
\label{R-matrix} \eea in which $E_{ij}$ is the matrix with
elements $(E_{ij})^l_k=\d_{jk}\d_{il}$. The coefficient functions
are \bea &&R^{ii}_{ii}(z,\l)=1,~~
R^{ij}_{ij}(z,\l)=\frac{\s(z)\s(\l_{ij}w+w)}
{\s(z+w)\s(\l_{ij}w)},\label{Elements1}\\
&& R^{ji}_{ij}(z,\l)=\frac{\s(w)\s(z+\l_{ij}w)}
{\s(z+w)\s(\l_{ij}w)},
\label{Elements2}
\eea
and  $\l_{ij}$ is
defined in (\ref{Def1}). The R-matrix satisfies the dynamical
(modified) quantum Yang-Baxter equation
\bea
&&R_{12}(z_1-z_2,\l-h^{(3)})R_{13}(z_1-z_3,\l)
R_{23}(z_2-z_3,\l-h^{(1)})\no\\
&&~~~~=R_{23}(z_2-z_3,\l)R_{13}(z_1-z_3,\l-h^{(2)})R_{12}(z_1-z_2,\l),
\label{MYBE}
\eea
with the initial condition
\bea
R^{kl}_{ij}(0,\l)=\d^l_i~\d^k_j.\label{Initial}
\eea
We adopt the
notation: $R_{12}(z,\l-h^{(3)})$ acts on a tensor $v_1\otimes v_2
\otimes v_3$ as $R(z;\l-\mu)\otimes Id$ if $v_3\in W[\mu]$.

A representation of the elliptic quantum group $\E$ (an
$\E$-module) is  by definition a pair ($W,L$) where $W$ is a
diagonalizable $\h$-module and $L(z,\l)$ is a meromorphic function
of $\l$ and the spectral parameter $z\in\Cb$,  with values in
$End_{\h}(\Cb^n\otimes W)$ (the endomorphism commuting with the
action of $\h$). It obeys the so-called $``RLL"$ relation
\bea
&&R_{12}(z_1-z_2,\l-h^{(3)})L_{13}(z_1,\l)
L_{23}(z_2,\l-h^{(1)})\no\\
&&~~~~=L_{23}(z_2,\l)L_{13}(z_1,\l-h^{(2)})R_{12}(z_1-z_2,\l),
\label{Exchange}
\eea
where the first and second space are auxiliary  spaces
($\Cb^n$) and the third space plays the role of quantum space
($W$). The total weight conservation condition for the
$L$-operator reads
\bea [h^{(1)}+h^{(3)},~L_{13}(z,\l)]=0.\no
\eea
In terms of the elements of the $L$-operator defined by
\bea
L(z,\l)\lt(e_i\otimes v\rt)=\sum_{j=1}^n e_j\otimes
L^j_i(z,\l)v,~~v\in W,
\eea the above condition can be expressed
equivalently as
\bea
f(h)L_i^j(z,\l)=L^j_i(z,\l)f(h+\hat{i}-\hat{j}),\label{Charge}
\eea
in which $f(h)$ is any meromorphic function of $h$ and $h$
measures the weight of the quantum space ($W$).

\subsection{Modules over $\E$ and the associated operator algebra}
The basic example of an $\E$-module is $(\Cb^n, L)$ with
$L(z,\l)=R(z-z_1,\l)$, which is called  the fundamental vector
representation $V_{\L_1}(z_1)$ with the evaluation point $z_1$. It
is obvious that $``RLL"$ relation is satisfied as a consequence of
the dynamical Yang-Baxter equation (\ref{MYBE}). Other modules can
be obtained by taking tensor products: if $(W_1,L^{(1)})$ and
$(W_2,L^{(2)})$ are $\E$-modules, where $L^{(j)}$ acts on
($\Cb^n\otimes W_j$), then also $(W_1\otimes W_2, L)$ with
\bea
L(z,\l)=L^{(1)}(z,\l-h^{(2)})L^{(2)}(z,\l)~~{\rm acting~ on}
~\Cb^n\otimes W_1\otimes W_2.\label{Fusion}
\eea

An $\E$-submodule of an $\E$-module $(W,L)$ is a pair $(W_1,L_1)$
where $W_1$ is an $\h$-submodule of $W$ such that $\Cb^n\otimes
W_1$ is invariant under the action of all the $L(z,\l)$, and
$L_1(z,\l)$ is the restriction to this invariant subspace. Namely,
the $\E$-submodules are $\E$-modules.

Using the fusion rule of $\E$ (\ref{Fusion}) one can construct the
symmetric $\E$-submodule of $l$-tensors of fundamental vector
representations: \bea V_{\L^{(l)}}(z_1)=~symmetric~
subspace~of~V_{\L_1}(z_1)\otimes
V_{\L_1}(z_1-w)\otimes\cdots\otimes V_{\L_1}(z_1-(l-1)w),\no \eea
where $\L^{(l)}$ is defined by (\ref{Weight}). We call such an
$\E$-module  the {\it higher spin-$l$ representation} with
evaluation point $z_1$. These series of representations in the
case of $\Zb_n$ Sklyanin algebra have been studied in \cite{Skl83}
for $n=2$ case and in \cite{Hou97,Has97} for generic $n$ case.
 From direct calculation, we find  that the $\E$-module
$V_{\L^{(l)}}(z)$ is an irreducible highest weight module of $\E$
with the highest vector $|\L^{(l)}\rangle\in W[l\L_1]=\Cb
|\L^{(l)}\rangle$. It satisfies the following highest weight
conditions
\bea
&&L^1_1(u,\l)|\L^{(l)}\rangle=|\L^{(l)}\rangle,~~L^i_1(u,\l)|\L^{(l)}
\rangle=0,~~i=2,\cdots,n,\\
&&L^{i}_{j}(u,\l)|\L^{(l)}\rangle=\d^{i}_{j} \frac
{\s(u-z)~\s(\l_{i1}w+lw)} {\s(u-z+lw)~\s(\l_{i1}w)}
|\L^{(l)}\rangle,~~i,j=2,\cdots,n,\\
&&f(h)|\L^{(l)}\rangle=f(l\hat{1})\,|\L^{(l)}\rangle,
\eea
where $f(h)$ is any meromorphic function of $h$, which measures
the weight of the quantum space $W$.

For any $\E$-module, as in \cite{Fel96} one can define an
associated operator algebra of difference operators on the space
$Fun(W)$ of meromorphic functions of $\l\in \mathbb{P}_{\x}$ with
values in $W$. The algebra is  generated by $h$ and the operators
$\tilde{L}(z)\in End(\Cb^n\otimes Fun(W))$ acting as
\bea
\tilde{L}(z)(e_i\otimes f)(\l)=\sum_{j=1}^{n}e_j\otimes
L_i^{j}(z,\l)f(\l-\hat{i}).\label{Definition2}
\eea
One can derive  the following
exchange relation of the difference operator $\tilde{L}(z)$
from the ``RLL" relation (\ref{Exchange}), the weight
conservation condition of $L_i^j(z,\lambda)$ (\ref{Charge})
  and  the fact
  $[h^{(1)}+h^{(2)},~R_{12}(z,\l)]=0$,
\bea
&&R_{12}(z_1-z_2,\l-h)\tilde{L}_{13}(z_1) \tilde{L}_{23}(z_2)
=\tilde{L}_{23}(z_2)\tilde{L}_{13}(z_1)R_{12}(z_1-z_2,\l),
\label{Exchange1}\\
&&f(h)\tilde{L}_i^j(z)=\tilde{
L}^j_i(z)f(h+\hat{i}-\hat{j}),\label{Exchange2}
\eea
where $f(h)$
is any meromorphic function of $h$ and $h$ measures the weight of
the quantum space $W$.

The {\it transfer matrices} associated to an $\E$-module $(W, L)$
\cite{Fel96} is a difference operator acting on the space
$Fun(W)[0]$ of meromorphic functions of $\l$ with values in the
zero-weight space of $W$. It is defined by
\bea
T(u)f(\l)=\sum_{i=1}^{n}\tilde{L}^i_i(u)f(\l)
=\sum_{i=1}^{n}L^{i}_{i}(u,\l)f(\l-\hat{i}). \label{transfer}
\eea
The exchange relation of $\tilde{L}$-operators
(\ref{Exchange1}) and (\ref{Exchange2}) imply that, for any
$\E$-module,  the above transfer matrices  %defined by (\ref{transfer})
preserve the space $H=Fun(W)[0]$ of functions with values in the
zero weight space $W[0]$. Moreover,  they commute pairwise on $H$:
\[
\left.[T(u)~,~T(v)]\vTm\right|_{H}=0.
\]

In  this paper we will study  the tensor product $\E$-module $W=
V_{\L^{(l_1)}}(z_1)\otimes V_{\L^{(l_2)}}(z_2)\otimes
\cdots\otimes V_{\L^{(l_m)}}(z_m)$. With the generic evaluation
points $\lt\{z_i\rt\}$, the module is an irreducible highest weight
$\E$-module \cite{Fel94,Fel961}. Let
$\L=\L^{(l_1)}+\cdots+\L^{(l_m)}$, then $W[\L]=\Cb |\L\rangle$
with
\bea
|\L\rangle\equiv|\L^{(l_1)},\L^{(l_2)},\cdots,
\L^{(l_m)}\rangle\equiv
|\L^{(l_1)}\rangle\otimes|\L^{(l_2)}\rangle\otimes\cdots\otimes
|\L^{(l_m)}\rangle.\label{hweight}
\eea
The vector $|\L\rangle$,
viewed as a constant function in $Fun(W)$, obeys the following
highest weight conditions:
\bea
&&\tilde{L}^1_1(z)|\L\rangle=A(z,\l)|\L\rangle,~~\tilde{L}^i_1(z)|\L
\rangle=0,~~i=2,\cdots,n,\no\\
&&\tilde{L}^{i}_{j}(z)|\L\rangle=\d^{i}_{j}
D_i(z,\l)|\L\rangle,~~i,j=2,\cdots,n,~~~
f(h)|\L\rangle=f(N\hat{1})|\L\rangle.\no \eea The highest weight
functions read \bea A(z,\l)=1,~~
D_i(z,\l)=\lt\{\prod_{k=1}^{m}\frac{\s(z-p_k)}{\s(z-q_k)}\rt\}
\frac{\s(\l_{i1}w+Nw)}{\s(\l_{i1}w)},~~i=2,\cdots,n,\label{Func}
\eea where \bea p_k=z_k,~~q_k=z_k-l_kw,~~~
N=\sum_{k=1}^{m}l_k,~~k=1,\cdots,m.\label{number} \eea

%%%%%%%%%%%%%%%%%%%%%%%%%%%%%%%%%%%%%%%%%%%%%%%%%%%%%%%%%%%%%%%
%                                                             %
%  3. Algebraic Bethe ansatz for $\E$                         %
%                                                             %
%%%%%%%%%%%%%%%%%%%%%%%%%%%%%%%%%%%%%%%%%%%%%%%%%%%%%%%%%%%%%%%
\section{Algebraic Bethe ansatz for $\E$}
\label{BA} \setcounter{equation}{0} In this section we fix a
highest weight $\E$-module $W$ of weight $\L$, the functions
$A(z,\l), D_i(z,\l)$  (\ref{Func}), with the highest vector
$|\L\rangle$. We assume that $N=\sum_{k=1}^{m}l_k=n\times l$ with
$l$ being an integer, so that the zero-weight space $W[0]$ can be
non-trivial and so that the algebraic Bethe ansatz method can be
applied as in \cite{Bax73, Hou89, Tak92, Fel96}.

Let us adopt the standard notation for convenience:
\bea
&&\A(u)=\tilde{L}^1_1(u),~~\B_i(u)=\tilde{L}^1_i(u),~~i=2,
\cdots,n,\label{Con1}\\
&&\C_i(u)=\tilde{L}^i_1(u),~~~\D^{j}_{i}(u)=\tilde{L}^j_i(u),
~~i,j=2,\cdots,n.\label{Con2}
\eea
The transfer matrices $T(u)$
become
\bea
T(u)=\A(u)+\sum_{i=2}^{n}\D^i_i(u).\label{transfer1}
\eea
Any non-zero vector $|\O\rangle\in Fun(W)[\L]$ is of form
$|\O\rangle=g(\l)|\L\rangle$, for some meromorphic function $g\ne
0$. When $N=n\times l$, the weight $\L$ can be  written in the
form
\bea \L=nl\L_1=l\sum_{k=1}^{n-1}(\e_1-\e_{k+1}).
\eea
Noting
(\ref{Exchange2}), the zero-weight vector space is spanned by the
vectors of the following form
\bea
  \B_{i_{N_1}}(v_{N_1})\B_{i_{N_1-1}}(v_{N_1-1})\cdots\B_{i_1}(v_1)
|\O\rangle, \label{Zero}
\eea
where $N_1=(n-1)\times l$ and among
the indices $\lt\{i_k~|~k=1,\cdots,N_1\rt\}$, the number of
$i_k=j$, denoted
  by $\#(j)$, should be
\bea
  \#(j)=l, ~~~j=2, \cdots, n.\label{zero-weight}
\eea
The above  states (\ref{Zero})  actually  belong to the
zero-weight space $W[0]$, because
\bea
&&f(h)\B_{i_{N_1}}(v_{N_1})\B_{i_{N_1-1}}(v_{N_1-1})\cdots\B_{i_1}(v_1)
|\O\rangle\no\\
&&~~~=\B_{i_{N_1}}(v_{N_1})\B_{i_{N_1-1}}(v_{N_1-1})\cdots\B_{i_1}(v_1)
f(h+\sum_{k=1}^{N_1}\hat{i}_k-N_1\hat{1})|\O\rangle\no\\
&&~~~=f(l\sum_{i=2}^{n}\hat{i}-(n-1)l\hat{1}+nl\hat{1})
\B_{i_{N_1}}(v_{N_1})\B_{i_{N_1-1}}(v_{N_1-1})\cdots\B_{i_1}(v_1)
|\O\rangle\no\\
&&~~~=f(0)
\B_{i_{N_1}}(v_{N_1})\B_{i_{N_1-1}}(v_{N_1-1})\cdots\B_{i_1}(v_1)
|\O\rangle,\no \eea for any meromorphic function $f$. We will seek
the common eigenvectors of the {\it transfer matrices} $T(u)$ in
the form \bea |\l;\lt\{v_k\rt\}\rangle
=F^{i_1,i_2,\cdots,i_{N_1}}(\l;\lt\{v_k\rt\})
\B_{i_{N_1}}(v_{N_1})\B_{i_{N_1-1}}(v_{N_1-1})
\cdots\B_{i_1}(v_1)|\O\rangle,\label{Eigenstate} \eea with the
restriction condition (\ref{zero-weight}) and the parameters
$\lt\{v_k\rt\}$ will be specified later by the Bethe ansatz
equations (\ref{BAE}). We adopt here  the convention that the
repeated indices imply summation over $2,3,\cdots,n$, and the
notation that \bea v_k=v^{(0)}_k,~k=1,2,\cdots, N_1. \eea For
convenience, let us introduce the following set of integers  \bea
N_i=(n-i)\times l,~~i=1,2,\cdots, n-1, \eea and $(n-1)$ complex
parameters $\{\a^{(i)}|i=1,\cdots,n-1\}$ and set
$\a^{(n)}=-\sum_{k=1}^{n-1}\a^{(k)}$ to specify quasi-vacuums of
each step of the nested Bethe ansatz (see below). Associated with
$\{\a^{(i)}\}$, set \bea
\bar{\alpha}^{(i)}=\frac{1}{(n-i-1)}\lt\{\alpha^{(i+1)}-\frac
{\sum_{k=i+1}^n\alpha^{(k)}}{n-i}\rt\}, ~~i=0,\cdots,
n-2.\label{Relation} \eea

 From the exchange relations (\ref{Exchange1}) and
(\ref{Exchange2}), one can derive the commutation relations among
$\A(u),~\D^j_i(u)$ and $\B_i(u)$ $(i,j=2,\cdots,n)$ (for details,
see  Appendix A). The relevant commutation relations are
\bea
&&\A(u)\B_i(v)=r(v-u,\l_{i1})\B_i(v)\A(u)+s(u-v,\l_{1i})\B_i(u)A(v),
\label{Rev1}\\
&&\D^j_i(u)\B_l(v)=r(u-v,~\l_{j1}-h_{j1})\sum_{\a,\b=2}^{n}
R^{\a\b}_{i~l}(u-v,\l)\B_{\b}(v)\D^j_{\a}(u)\no\\
&&~~~~~~~~~~~~~~~~~~-s(u-v,\l_{1j}-h_{1j})\B_i(u)\D^j_l(v),\label{Rev2}\\
&&\B_i(u)\B_j(v)=\sum_{\a,\b=2}^{n}R^{\a\b}_{i~j}(u-v,\l)
\B_{\b}(v)\B_{\a}(u),\label{Rev3}\\
&&f(h)\A(u)=\A(u)f(h),~~f(h)\B_i(u)=\B_i(u)f(h+\hat{i}-\hat{1}),
\label{Rev4}\\
&&f(h)\D^j_i(u)=\D^j_i(u)f(h+\hat{i}-\hat{j}),\label{Rev5} \eea
where the function $r,s$ are defined by \bea
r(u,\nu)=\frac{\s(u+w)\s(\nu w)}{\s(u)\s(\nu w+w)},~~
s(u,\nu)=\frac{\s(w)\s(u+\nu w)}{\s(u)\s(\nu w-w)},~~\nu\in
\Cb.\no \eea Because of the simple poles in  the functions $r,~s$
at $u\in \Zb+\tau\Zb$, let us  assume for convenience that all
$\lt\{v_k\rt\}$ are distinct $modulo~ \Zb+\tau\Zb$ and consider
$u$ at a generic point. Take \bea
|\O\rangle=g(\l)|\L\rangle,\label{Vac-f} \eea as the so-called
quasi-vacuum satisfying the following conditions \bea
&&\A(u)|\O\rangle=\frac{g(\l-\hat{1})}{g(\l)}|\O\rangle,~~\C_i(u)|\O
\rangle=0,~~i=2,\cdots,n,\\
&&\D^{j}_{i}(u)|\O\rangle=\d^{i}_{j}D_i(u,\l)
~\frac{g(\l-\hat{i})}{g(\l)} |\O\rangle,~~i,j=2,\cdots,n,\\
&&f(h)|\O\rangle=f(N\hat{1})|\O\rangle, ~~\B_i(u)|\O\rangle\ne 0,
~~i=2,\cdots,n,
\eea
where $D_i(u,\l)$, $p_k$ and $q_k$ are given
in (\ref{Func}) and (\ref{number}). We will specify the function
$g(\l)$ later (\ref{g-fun})

Now, let us evaluate the action of $\A(u)$ on
$|\l;\lt\{v_k\rt\}\rangle$. Many terms will appear when we move
$\A(u)$ from the left  to the right  of
$\B_i(v_k)$'s. They can be classified into two types: wanted and
unwanted terms. The wanted terms in
$\A(u)|\l;\lt\{v_k\rt\}\rangle$ can be obtained by retaining the
first term in the commutation relation (\ref{Rev1}). The unwanted
terms arising from the second term of (\ref{Rev1}), have some
$B(v_k)$  replaced by $B(u)$. One unwanted term where $B(v_{N1})$
is replaced by $B(u)$ can be obtained by using firstly the second
terms of (\ref{Rev1}), then repeatedly using the first term of
(\ref{Rev1}). Thanks to the commutation relation (\ref{Rev3}), one
can easily obtain {\it the other unwanted terms} where $B(v_k)$ is
replaced by $B(u)$. So, we can find the action of $A(u)$ on
$|\l;\lt\{v_k\rt\}\rangle$
\bea
&&\A(u)~F^{i_1,i_2,\cdots,i_{N_1}}(\l;\lt\{v_k\rt\})
\B_{i_{N_1}}(v_{N_1})\B_{i_{N_1-1}}(v_{N_1-1})
\cdots\B_{i_1}(v_1)|\O\rangle\no\\
&&~=F^{i_1,i_2,\cdots,i_{N_1}}(\l-\hat{1};\lt\{v_k\rt\})
\lt\{\prod_{k=1}^{N_1}
r(v_k-u,\langle\l-\sum_{s=k+1}^{N_1}\hat{i}_s,\e_{i_k}-\e_1\rangle)\rt\}\no\\
&&~~~~\times B_{i_{N_1}}(v_{N_1})\cdots
B_{i_1}(v_1)\frac{g(\l-\hat{1})}{g(\l)}|\O\rangle \no\\
&&~~+F^{i_1,i_2,\cdots,i_{N_1}}(\l-\hat{1};\lt\{v_k\rt\})
s(u-v_{N_1},\l_{1i_{N_1}}) \lt\{\prod_{k=1}^{N_1-1}
r(v_k-v_{N_1},\langle\l-\sum_{s=k+1}^{N_1}\hat{i}_s,\e_{i_k}-\e_1\rangle)\rt\}\no\\
&&~~~~\times B_{i_{N_1}}(u)B_{i_{N_1-1}}(v_{N_1-1})\cdots
B_{i_1}(v_1)\frac{g(\l-\hat{1})}{g(\l)}|\O\rangle \no \\
&&~~+o.u.t.,
\eea
where $o.u.t.$ stands for  {\it the other
unwanted terms}. Due to the restriction condition
(\ref{zero-weight}) and $N_1=(n-1)\times l$, we have
\bea
&&\lt\{\prod_{k=1}^{N_1}
r(v_k-u,\langle\l-\sum_{s=k+1}^{N_1}\hat{i}_s,\e_{i_k}-\e_1\rangle)\rt\}=
\lt\{\prod_{k=1}^{N_1}\frac{\s(v_k-u+w)}{\s(v_k-u)}\rt\}\lt\{\prod_{j=2}^{n}
\frac{\s(\l_{j1}w-lw+w)}{\s(\l_{j1}w+w)}\rt\},\no\\
&&~~\label{R-g}\\
&&\sum_{k=1}^{N_1}\hat{i}_{k}=l\sum_{i=2}^{n}\hat{i}=-l\hat{1}.\label{Rest1}
\eea In order that the Bethe ansatz equation (\ref{BAE}) and the
eigenvalues (\ref{t-function}) should be independent of $\l$, one
needs to  choose (cf. \cite{Fel96}) \bea g(\l)=e^{\langle
\a^{(1)}\e_1,~\l w\rangle}\prod_{j=2}^{n}\lt\{
\prod_{k=1}^{l}\frac{\s(\l_{j1}w+kw)}{\s(w)}\rt\}.\label{g-fun}
\eea  The action of $\A(u)$ on $|\l;\lt\{v_k\rt\}\rangle$ becomes
\bea &&\A(u)~F^{i_1,i_2,\cdots,i_{N_1}}(\l;\lt\{v_k\rt\})
\B_{i_{N_1}}(v_{N_1})\B_{i_{N_1-1}}(v_{N_1-1})
\cdots\B_{i_1}(v_1)|\O\rangle\no\\
&&~= e^{\frac{1-n}{n}\a^{(1)}
w}\prod_{k=1}^{N_1}\frac{\s(v_k-u+w)}{\s(v_k-u)}
F^{i_1,i_2,\cdots,i_{N_1}}(\l-\hat{1};\lt\{v_k\rt\})
B_{i_{N_1}}(v_{N_1})\cdots
B_{i_1}(v_1)|\O\rangle \no\\
&&~~+s(u-v_{N_1},\l_{1i_{N_1}})
\frac{\s(\l_{i_{N_1}1}w+w)}{\s(\l_{i_{N_1}1}w)}
e^{\frac{1-n}{n}\a^{(1)}
w}\prod_{k=1}^{N_1-1}\frac{\s(v_k-v_{N_1}+w)}{\s(v_k-v_{N_1})}\no\\
&&~~~~~\times
F^{i_1,i_2,\cdots,i_{N_1}}(\l-\hat{1};\lt\{v_k\rt\})
B_{i_{N_1}}(u)B_{i_{N_1-1}}(v_{N_1-1})\cdots
B_{i_1}(v_1)|\O\rangle \no \\
&&~~+o.u.t..
\eea
Next, we evaluate the action of $\D^{i}_{i}(u)$
on $|\l;\lt\{v_k\rt\}\rangle$:
\bea
&&\D^i_i(u)~F^{i_1,i_2,\cdots,i_{N_1}}(\l;\lt\{v_k\rt\})
\B_{i_{N_1}}(v_{N_1})\B_{i_{N_1-1}}(v_{N_1-1})
\cdots\B_{i_1}(v_1)|\O\rangle\no\\
&&~~=\lt\{\prod_{k=1}^{N_1}
r(u-v_k,\langle\l-h-(N_1-k)\hat{1},\e_{i}-\e_1\rangle)\rt\}\no\\
&&~~~~\times\lt\{\prod_{k=1}^{m}\frac{\s(u-p_k)}{\s(u-q_k)}\rt\}
~\frac{\s(\l_{i1}w+(N-l)w)}{\s(\l_{i1}w-lw)}
~\frac{g(\l-\hat{i}+l\hat{1})}
  {g(\l+l\hat{1})}\no\\
  &&~~~~\times \lt\{
  {L^{(1)}}^i_i(u,\l)\rt\}
  ^{i_1',i_2',\cdots,i_{N_1}'}
  _{i_1,i_2,\cdots,i_{N_1}}~F^{i_1,i_2,\cdots,i_{N_1}}
  (\l-\hat{i};\lt\{v_k\rt\})\no\\
  &&~~~~\times
\B_{i_{N_1}'}(v_{N_1})\B_{i_{N_1-1}'}(v_{N_1-1})
\cdots\B_{i_1'}(v_1)|\O\rangle\no\\
&&~~+u.t.,
\eea
where {\it u.t.} stands for the all {\it unwanted
terms}. We have introduced the convenient notation
\bea
\lt\{{L^{(1)}}^j_i(u,\l)\rt\}
  ^{i_1',i_2',\cdots,i_{N_1}'}
  _{i_1,i_2,\cdots,i_{N_1}}
  &=& R^{j,~i_1'}_{\g_{N_1-1},~i_1}
  (u-v_1;\l-\sum_{k=2}^{N_1}\hat{i}_k')
  R^{\g_{N_1-1},~i_2'}_{\g_{N_1-2},~i_2}
  (u-v_2;\l-\sum_{k=3}^{N_1}\hat{i}_k')\no\\
  &&~~\times \cdots
  R^{\g_1,~i_{N_1}'}_{i,~i_{N_1}}
  (u-v_{N_1};\l),\label{Cnotation}
  \eea
and all the indices take the values $2,\cdots,n$. Noting the
explicit form of the function $g(\l)$ given in (\ref{g-fun}) and the
zero-weight of the state $|\l;\lt\{v_k\rt\}\rangle$, the action of
$\D^{i}_{i}(u)$ becomes
\bea
&&\D^i_i(u)|\l;\lt\{v_k\rt\}\rangle\no\\
&&~=e^{\frac{\a^{(1)}}{n} w}
\lt\{\prod_{k=1}^{N_1}\frac{\s(u-v_k+w)}{\s(u-v_k)}\rt\}
\lt\{\prod_{k=1}^{m}\frac{\s(u-p_k)}{\s(u-q_k)}\rt\}\no\\
&&~~~~~~\times \lt\{
  {L^{(1)}}^i_i(u,\l)\rt\}
  ^{i_1',i_2',\cdots,i_{N_1}'}
  _{i_1,i_2,\cdots,i_{N_1}}~F^{i_1,i_2,\cdots,i_{N_1}}
  (\l-\hat{i};\lt\{v_k\rt\})\no\\
  &&~~~~~~\times
\B_{i_{N_1}'}(v_{N_1})\B_{i_{N_1-1}'}(v_{N_1-1})
\cdots\B_{i_1'}(v_1)|\O\rangle\no\\
&&~~~-e^{\frac{\a^{(1)}}{n} w}s(u-v_{N_1},\l_{1i})
\frac{\s(\l_{i1}w+w)}{\s(\l_{i1}w)}
\lt\{\prod_{k=1}^{N_1-1}\frac{\s(v_{N_1}-v_k+w)}{\s(v_{N_1}-v_k)}\rt\}
\lt\{\prod_{k=1}^{m}\frac{\s(v_{N_1}-p_k)}{\s(v_{N_1}-q_k)}\rt\}\no\\
&&~~~~~~\times \lt\{
  {L^{(1)}}^i_i(v_{N_1},\l)\rt\}
  ^{i_1',i_2',\cdots,i_{N_1}'}
  _{i_1,i_2,\cdots,i_{N_1}}~F^{i_1,i_2,\cdots,i_{N_1}}
  (\l-\hat{i};\lt\{v_k\rt\})\no\\
  &&~~~~~~\times
\B_{i_{N_1}'}(u)\B_{i_{N_1-1}'}(v_{N_1-1})
\cdots\B_{i_1'}(v_1)|\O\rangle\no \\
&&~~~+o.u.t..
\eea
Keeping the initial condition of R-matrix
(\ref{Initial}) and the notation (\ref{Cnotation}) in mind, we can
find the action of the transfer matrices on the state
$|\l;\lt\{v_k\rt\}\rangle$:
\bea &&T(u)|\l;\lt\{v_k\rt\}\rangle=
\lt(\A(u)+\sum_{i=2}^n\D^i_i(u)\rt)|\l;\lt\{v_k\rt\}\rangle\no\\
&&~= e^{\frac{1-n}{n}\a^{(1)}
w}\prod_{k=1}^{N_1}\frac{\s(v_k-u+w)}{\s(v_k-u)}
F^{i_1,i_2,\cdots,i_{N_1}}(\l-\hat{1};\lt\{v_k\rt\})
B_{i_{N_1}}(v_{N_1})\cdots
B_{i_1}(v_1)|\O\rangle \no\\
&&~~~+e^{\frac{\a^{(1)}}{n} w}
\lt\{\prod_{k=1}^{N_1}\frac{\s(u-v_k+w)}{\s(u-v_k)}\rt\}
\lt\{\prod_{k=1}^{m}\frac{\s(u-p_k)}{\s(u-q_k)}\rt\}\no\\
&&~~~~~~\times \lt\{
  T^{(1)}(u)\rt\}
  ^{i_1',i_2',\cdots,i_{N_1}'}
  _{i_1,i_2,\cdots,i_{N_1}}~F^{i_1,i_2,\cdots,i_{N_1}}
  (\l;\lt\{v_k\rt\})\no\\
  &&~~~~~~\times
\B_{i_{N_1}'}(v_{N_1})\B_{i_{N_1-1}'}(v_{N_1-1})
\cdots\B_{i_1'}(v_1)|\O\rangle\no\\
&&~~~+s(u-v_{N_1},\l_{1i_{N_1}})
\frac{\s(\l_{i_{N_1}1}w+w)}{\s(\l_{i_{N_1}1}w)}
e^{\frac{1-n}{n}\a^{(1)}
w}\prod_{k=1}^{N_1-1}\frac{\s(v_k-v_{N_1}+w)}{\s(v_k-v_{N_1})}\no\\
&&~~~~~~\times
F^{i_1,i_2,\cdots,i_{N_1}}(\l-\hat{1};\lt\{v_k\rt\})
B_{i_{N_1}}(u)B_{i_{N_1-1}}(v_{N_1-1})\cdots
B_{i_1}(v_1)|\O\rangle \no \\
&&~~~-\lt(\sum_{i=2}^ne^{\frac{\a^{(1)}}{n} w}s(u-v_{N_1},\l_{1i})
\frac{\s(\l_{i1}w+w)}{\s(\l_{i1}w)}
\lt\{\prod_{k=1}^{N_1-1}\frac{\s(v_{N_1}-v_k+w)}{\s(v_{N_1}-v_k)}\rt\}
\lt\{\prod_{k=1}^{m}\frac{\s(v_{N_1}-p_k)}{\s(v_{N_1}-q_k)}\rt\}\rt.\no\\
&&~~~~~~~\times \d^i_{i_{N_1}'}\lt\{
  T^{(1)}(v_{N_1})\rt\}
  ^{i_1',i_2',\cdots,i_{N_1}'}
  _{i_1,i_2,\cdots,i_{N_1}}~F^{i_1,i_2,\cdots,i_{N_1}}
  (\l;\lt\{v_k\rt\})\no\\
  &&~~~~~~\times \lt.
\B_{i_{N_1}'}(u)\B_{i_{N_1-1}'}(v_{N_1-1})
\cdots\B_{i_1'}(v_1)|\O\rangle\rt)\no \\
&&~~~+o.u.t., \label{Main}
\eea
where we have introduced the
reduced transfer matrices $T^{(1)}(u)$
\bea T^{(1)}(u)
F^{i_1',i_2',\cdots,i_{N_1}'}
  (\l;\lt\{v_k\rt\})=
\sum_{i=2}^n {L^{(1)}}^i_i(u,\l)
  ^{i_1',i_2',\cdots,i_{N_1}'}
  _{i_1,i_2,\cdots,i_{N_1}}~
  F^{i_1,i_2,\cdots,i_{N_1}}
  (\l-\hat{i};\lt\{v_k\rt\}).\no\\
  \label{reduce-transfer}
\eea Thanks to  the commutation relations (\ref{Rev3}), one can
obtain the explicit expressions of all the {\it o.u.t}. The
equation (\ref{Main}) tells that the state $
|\l;\lt\{v_k\rt\}\rangle$ is not an eigenvector of the transfer
matrices $T(u)$ {\em unless\/} $F's$ are the eigenvectors of the
reduced transfer matrices $T^{(1)}(u)$. The condition that the
third and fourth terms in the above equation should cancel each
other and also the all $o.u.t.$ terms vanish, will give a
restriction on the $N_1$ parameters $\lt\{v_k\rt\}$, the so-called
Bethe ansatz equations. Hence we arrive at the final results: \bea
T(u)|\l;\lt\{v_k\rt\}\rangle=t(u;\lt\{v_k\rt\}) |\l;\lt\{v_k\rt\}
\rangle. \eea The eigenvalue reads \bea t(u;\lt\{v_k\rt\})&=&
e^{(1-n)\bar{\a} w}
\lt\{\prod_{k=1}^{N_1}\frac{\s(v_k-u+w)}{\s(v_k-u)}\rt\}\no\\
&&~~~+e^{\bar{\a} w}
\lt\{\prod_{k=1}^{N_1}\frac{\s(u-v_k+w)}{\s(u-v_k)}\rt\}
\lt\{\prod_{k=1}^{m}\frac{\s(u-p_k)}{\s(u-q_k)}\rt\}
t^{(1)}(u;\{v^{(1)}_k\}),\no\\
&~~\label{t-function}
\eea
in which
\bea
&&T^{(1)}(u)
F^{i_1',i_2',\cdots,i_{N_1}'}
  (\l;\lt\{v_k\rt\})=e^{-\frac{\sum_{i=2}^{n}\a^{(i)}}{n(n-1)}w}
  t^{(1)}(u;\{v^{(1)}_k\})F^{i_1',i_2',\cdots,i_{N_1}'}
  (\l;\lt\{v_k\rt\}),\label{Re-P}\\
&&F^{i_1',i_2',\cdots,i_{N_1}'}(\l-\hat{1};\lt\{v_k\rt\})=
e^{\frac{\sum_{i=2}^n\a^{(i)}}{n}w} F^{i_1',i_2',\cdots,i_{N_1}'}
(\l;\lt\{v_k\rt\}), \eea and $\lt\{ v_k~|~k=1,2,\cdots, N_1\rt\}$
satisfy \bea e^{-n\bar{\a} w} \lt\{ \prod_{k=1,k\ne
s}^{N_1}\frac{\s(v_k-v_s+w)}{\s(v_k-v_s-w)}\rt\}= \lt\{
\prod_{k=1}^{m}\frac{\s(v_s-p_k)}{\s(v_s-q_k)}\rt\}
t^{(1)}(v_s;\{v^{(1)}_k\})\label{BA1}. \eea The parameters
$\{v^{(1)}_k|~k=1,2,\cdots, N_2\}$ will be specified later by the
Bethe ansatz equations (\ref{BAE}).

The diagonalization of the transfer matrices $T(u)$  is now
reduced to diagonalization of the reduced transfer matrices
$T^{(1)}(u)$ in (\ref{Re-P}). The explicit expression of
$T^{(1)}(u)$ given in (\ref{reduce-transfer}) implies that
$T^{(1)}(u)$ can be considered as the transfer matrices of an
$\r$-module $W^{(1)}$ (or reduced space): $N_1$ tensor product of
fundamental representations of $\r$ with evaluation points
$\lt\{v_k\rt\}$. We can use the same method to find the eigenvalue
of $T^{(1)}(u)$ as we have done for the diagonalization of $T(u)$.
Similarly to (\ref{Con1})---(\ref{Con2}), we introduce \bea
&&\A^{(1)}(u)=\{\tilde{L}^{(1)}\}^2_2(u),~~\B^{(1)}_i(u)=
\{\tilde{L}^{(1)}\}^2_i(u),~~i=3, \cdots,n,\\
&&\C^{(1)}_i(u)=\{\tilde{L}^{(1)}\}^i_2(u),~~~{\D^{(1)}}^{j}_{i}(u)
=\{\tilde{L}^{(1)}\}^j_i(u), ~~i,j=3,\cdots,n. \eea Then the
reduced transfer matrices (\ref{reduce-transfer})  can be
rewritten \bea
T^{(1)}(u)=\A^{(1)}(u)+\sum_{i=3}^n{\D^{(1)}}^i_i(u). \eea We seek
the common eigenvectors of the reduced transfer matrices
$T^{(1)}(u)$ in an analogous form  to (\ref{Eigenstate}): \bea
|\l;\{v^{(1)}_k\}\rangle^{(1)}
=\hspace{-7pt}\sum_{i_1,\cdots,i_{N_2}=3}^n\hspace{-7pt}
{F^{(1)}}^{i_1,i_2,\cdots,i_{N_2}}(\l;\{v^{(1)}_k\})
\B^{(1)}_{i_{N_2}}(v_{N_2}^{(1)})\B^{(1)}_{i_{N_2-1}}(v_{N_2-1}^{(1)})
\cdots\B^{(1)}_{i_1}(v_1^{(1)})|\O^{(1)}\rangle,\no\\
\label{Eigenstate1}
\eea
with the restriction of zero-weight
conditions similar to (\ref{zero-weight}).
The quasi-vacuum $|\O^{(1)}\rangle$
is the corresponding highest weight vector in the reduced space
$W^{(1)}$. We can also derive the relevant commutation
relations among $\A^{(1)}(u)$, ${\D^{(1)}}^j_i$ and $B_i^{(1)}(u)$
$(i,j=3,4,\cdots,n)$ similar  to (\ref{Rev1})---(\ref{Rev5}). Using
the same method as we have done for the diagonalization of $T(u)$,
we obtain the eigenvalue of the reduced transfer matrices
$T^{(1)}(u)$:
\bea
T^{(1)}(u)|\l;\{v^{(1)}_k\}\rangle^{(1)}=
t^{(1)}(u;\{v^{(1)}_k\}) |\l;\{v^{(1)}_k\}
\rangle^{(1)}.
\eea
Though very complicated, the coefficients
$F^{i_1,i_2,\cdots,i_{N_1}}(\l;\{v_k\})$ could be extracted from the
eigenvectors $|\l;\{v^{(1)}_k\}\rangle^{(1)}$,  in principle.
  The eigenvalue
$t^{(1)}(u;\{v^{(1)}_k\})$ is given by \bea
t^{(1)}(u;\{v^{(1)}_k\})&=& e^{(2-n)\bar{\a}^{(1)} w}
\lt\{\prod_{k=1}^{N_2}\frac{\s(v^{(1)}_k-u+w)}
{\s(v^{(1)}_k-u)}\rt\}\no\\
&&~~~+e^{\bar{\a}^{(1)} w}
\lt\{\prod_{k=1}^{N_2}\frac{\s(u-v^{(1)}_k+w)}
{\s(u-v^{(1)}_k)}\rt\}
\lt\{\prod_{k=1}^{N_1}\frac{\s(u-v_k)}{\s(u-v_k+w)}\rt\}
t^{(2)}(u;\{v^{(2)}_k\}).\no\\
&~~\label{t-function1}
\eea
The parameters
$\{v^{(2)}_k|~k=1,2,\cdots,N_3\}$ will be specified later by the
Bethe ansatz equations (\ref{BAE}). The function
$t^{(2)}(u;\{v^{(2)}_k\})$ is the eigenvalue of the second
reduced transfer matrices $T^{(2)}(u)$
\bea
&&T^{(2)}(u)
{F^{(1)}}^{i_1',\cdots,i_{N_2}'}
  (\l;\{v^{(1)}_k\})=e^{-\frac{\sum_{i=3}^n\a^{(i)}}{(n-1)(n-2)}w}
  t^{(2)}(u;\{v^{(2)}_k\}){F^{(1)}}^{i_1',\cdots,i_{N_2}'}
  (\l;\{v^{(1)}_k\}),\label{Re-P1}
\eea and $T^{(2)}(u)$ is given  by (\ref{Cnotation}) and
(\ref{reduce-transfer}) with  all indices taking values over
$3,\cdots,n$ and  depending on  $\{v^{(1)}_k\}$ instead of
$\lt\{v_k\rt\}$. Hence, the diagonalization of $T(u)$ is further
reduced to the diagonalization of $T^{(2)}(u)$ in (\ref{Re-P1}).
This is so-called nested Bethe ansatz.  Repeating the above
procedure further $n-2$ times, one can reduce to the last reduced
transfer matrices $T^{(n-1)}(u)$ which is trivial to get the
eigenvalues. At the same time we need to introduce the
$\frac{n(n-1)}{2}l$ parameters $\{\{
v_k^{(i)}|~k=1,2,\cdots,N_{i+1}\},~i=0,1,\cdots,n-2\}$ to specify
the eigenvectors of the corresponding reduced transfer matrices
$T^{(i)}(u)$  (including the original one $T(u)=T^{(0)}(u)$), and
$n$ arbitrary complex numbers $\lt\{\a^{(i)}|i=1\cdots,n\rt\}$ to
specify the quasi-vacuum $|\O^{(i)}\rangle$ of each step as in
(\ref{Vac-f}) and (\ref{g-fun}) . Finally, we obtain all the
eigenvalues of the reduced transfer matrices $T^{(i)}(u)$ with the
eigenvalue $t^{(i)}(u;\{v^{(i)}_{k}\})$ in a recurrence form \bea
&&t^{(i)}(u;\{v^{(i)}_k\})= e^{(i+1-n)\bar{\a}^{(i)} w}
\lt\{\prod_{k=1}^{N_{i+1}}\frac{\s(v^{(i)}_k-u+w)}{\s(v^{(i)}_k-u)}\rt\}\no\\
&&~~~~~~+e^{\bar{\a}^{(i)} w}
\lt\{\prod_{k=1}^{N_{i+1}}\frac{\s(u-v^{(i)}_k+w)}{\s(u-v^{(i)}_k)}\rt\}
\lt\{\prod_{k=1}^{N_i}\frac{\s(u-p^{(i)}_k)}{\s(u-q^{(i)}_k)}\rt\}
t^{(i+1)}(u;\{v^{(i+1)}_k\}),\no\\[4pt]
&&~~~~~~~~~~~~~~~~~~~~~~~~~~~~~~~~i=0,1,\cdots,n-2,\\
&&t^{(0)}(u;\{v^{(0)}_k\}) =t(u;\lt\{v_k\rt\}),~~ t^{(n-1)}(u)=1,
\eea where $\bar{\a}^{(i)}$, $i=0,1,\cdots, n-2$ are given by
(\ref{Relation}), $\bar{\a}=\bar{\a^{(0)}}$, $N_0=m$ and \bea
&&p^{(0)}_k=p_k=z_k,~~q^{(0)}_k=q_k=z_k-l_kw,~~k=1,2,\cdots,m,
\\
&&p^{(i)}_k=v^{(i-1)}_k,~~q^{(i)}_k=v^{(i-1)}_k-w,~~i=1,2,\cdots,
n-2,~~~k=1,2,\cdots,N_{i}. \eea The $\{\{v^{(i)}_k\}\}$ satisfy
the following Bethe ansatz equations \bea e^{(i-n)\bar{\a}^{(i)}
w} \lt\{ \prod_{k=1,k\ne
s}^{N_{i+1}}\frac{\s(v^{(i)}_k-v^{(i)}_s+w)}{\s(v^{(i)}_k-v^{(i)}_s-w)}\rt\}=
\lt\{
\prod_{k=1}^{N_i}\frac{\s(v^{(i)}_s-p^{(i)}_k)}{\s(v^{(i)}_s-q^{(i)}_k)}\rt\}
t^{(i+1)}(v^{(i)}_s;\{v^{(i+1)}_k\}).\label{BAE} \eea

We conclude this section with some remarks on analytic
properties of the functions $t^{(i)}(u;\{v_k^{(i)}\})$. By
construction, the eigenvalue functions
$t^{(i)}(u;\{v_k^{(i)}\})$ of the transfer matrices
should not be singular at $u=v^{(i)}_k$
($modulo~\Zb+\tau\Zb$) with $0\leq i\leq n-2,~1\leq k\leq
N_{i+1}$. On the other hand, the Bethe ansatz equations (\ref{BA1}) and
(\ref{BAE}) are derived from the requirement that the unwanted terms
should vanish. It is interesting to note that
these constraints could be understood from a different point of view:
from equation
(\ref{t-function}), we know that $u=v^{(0)}_k=v_k$
($modulo~\Zb+\tau\Zb$)  is a possible simple pole position of
$t(u;\lt\{v_k\rt\})$.
However, the constraints  on $\{v_k\}$,
the Bethe ansatz equations (\ref{BA1})
simply tell that  the {\it residue} of $t(u;\lt\{v_k\rt\})$ at $u=v_k$
($modulo~\Zb+\tau\Zb$) is vanishing.
  Hence, $t(u;\lt\{v_k\rt\})$ is analytic at
$u=v_k$ ($modulo~\Zb+\tau\Zb$). Similarly, the Bethe ansatz
equations (\ref{BAE}) ensure the analyticity of
$t^{(i)}(u;\{v_k^{(i)}\})$ at all $u=v^{(i)}_k$ ($modulo~\Zb+\tau\Zb$).
Therefore, the eigenvalues of transfer matrices $T(u)$ are analytic
functions of $u$.
%%%%%%%%%%%%%%%%%%%%%%%%%%%%%%%%%%%%%%%%%%%%%%%%%%%%%%%%%%%%%%%%%%%%%
%                                                                   %
%The Z_n Belavin model                                              %
%                                                                   %
%%%%%%%%%%%%%%%%%%%%%%%%%%%%%%%%%%%%%%%%%%%%%%%%%%%%%%%%%%%%%%%%%%%%%
\section{The $\Zb_n$ Belavin model}
\label{Belavin} \setcounter{equation}{0} We shall show in this
section how to obtain the  eigenvectors and the corresponding
eigenvalues of the transfer matrices for  $\Zb_n$ Belavin model
with periodic boundary condition from our results in section 3.

The $\Zb_n$ Belavin model \cite{Bel81} is based on the following
R-matrix \bea R_B(u)=\sum_{i,j,k,l}R^{kl}_{ij}(u)E_{ik}\otimes
E_{lj}.\label{Belavin-R}\eea The coefficient functions are
\cite{Jim87} \bea R^{kl}_{ij}(u)=\lt\{
\begin{array}{ll}
\frac{h(u)\s(w)\t^{(i-j)}(u+w)}{\s(u+w)\t^{(i-k)}(w)\t^{(k-j)}(u)}&{\rm
if}~i+j=k+l~mod~n\\
0&{\rm otherwise}
\end{array}\rt.
\eea where \bea h(u)=\frac{\prod_{j=0}^{n-1}\t^{(j)}(u)}
{\prod_{j=1}^{n-1}\t^{(j)}(0)}.\eea The R-matrix satisfies the
quantum Yang-Baxter equation \bea
R_{12}(u_1-u_2)R_{13}(u_1-u_3)R_{23}(u_2-u_3)=
R_{23}(u_2-u_3)R_{13}(u_1-u_3)R_{12}(u_1-u_2),\label{YBE} \eea and
the initial condition: $R^{kl}_{ij}(0)=\d^l_i\d^k_j$. For generic
evaluation points $z_1,\cdots,z_m$, one then defines the commuting
transfer matrices \bea
&&T_{B}(u)=tr_{0}L_{B}(u),\label{T}\\
&&L_{B}(u)=\{R_B\}_{01}(u-z_1)\cdots
\{R_B\}_{0m}(u-z_m),\label{L}
\eea
where the L-operator $L_{B}(u)$
acts on $\Cb^n\otimes\lt(\Cb^n\rt)^{\otimes m}$ and the transfer
matrices $T_{B}(u)$ act on $\lt(\Cb^n\rt)^{\otimes m}$.

Using the intertwiner introduced by Jimbo et al \cite{Jim87}, we
can define the matrix $S(z,\l)$ with the  elements
$S^i_j(z,\l)=\frac{\t^{(i)}(z+nw\l_j)}{\prod_{k\neq
j}\s(\l_{kj}w)}$. When $\l\in \mathbb{P}_{\x}$, the inverse matrix
of $S(z,\l)$ exists. Then, one finds the following {\it twisting
relation\/} holds: \bea &&\{R_B\}_{12}(u_1-u_2)
S(u_1,\l)^{(1)} S(u_2,\l-h^{(1)})^{(2)}\no\\
&&~~~~~~~~=S(u_2,\l)^{(2)}S(u_1,\l-h^{(2)})^{(1)}
R_{12}(u_1-u_2,\l),\label{twisting}
\eea
in which $R(u,\lambda)$ is defined by (\ref{R-matrix}), (\ref{Elements1})
and (\ref{Elements2}).
The modified quantum
Yang-Baxter equation (\ref{MYBE}) and the quantum Yang-Baxter
equation (\ref{YBE}) are equivalent to each other due to the
above {\it
twisting relation} (\ref{twisting}). Let us introduce
\bea
S_{m}(z_1,\cdots,z_m;\l)=S(z_m,\l)^{(m)}
S(z_{m-1},\l-h^{(m)})^{(m-1)}\cdots
S(z_1,\l-\sum_{k=2}^{m}h^{(k)})^{(1)}.\label{Twistor}
\eea

For the special case of  $l_1=l_2=\cdots=l_m=1$,
%by iterating (\ref{twisting}),
we find that the L-operator
$L(u,\l)$ corresponding to the $\E$-module $W$ is equivalent to
the L-operator $L_B(u)$ of $\Zb_n$ Belavin model (\ref{L}) by the
{\it twisting relation}
\bea
&&S_{m}(z_1,\cdots,z_m;\l)S(u,\l-h)^{(0)}L(u,\l)\no\\
&&~~~~~~=L_{B}(u)
S(u,\l)^{(0)}S_{m}(z_1,\cdots,z_m;\l-h^{(0)}).\label{Twisting}
\eea
In this formula $S_m(z_1,\cdots,z_m;\l)$ acts on the factors from
$1$ to $m$, and $h=h^{(1)}+\cdots+h^{(m)}$. Acting the both sides
of (\ref{Twisting}) on a state $e_i\otimes v$ with $v\in W[0]$,
replacing $\l$ by $\l+\hat{i}$, and noting (\ref{Charge}), we find
\bea
&&\sum_{j=1}^n\lt(S(u,\l+\hat{j})e_j\rt)\otimes
\lt(S_m(z_1,\cdots,z_m;\l+\hat{i})L^j_i(u,\l+\hat{i})v\rt)\no\\
&&~~~~~~=L_B(u)\lt\{\lt(S(u,\l+\hat{i})e_i\rt)\otimes
\lt(S_m(z_1,\cdots,z_m;\l)v\rt)\rt\}.
\eea
Therefore,  one can
derive
\bea T_B(u)S_m(z_1,\cdots,z_m;\l)|_{H}=\sum_{i=1}^{n}
S_m(z_1,\cdots,z_m;\l+\hat{i})L^i_i(u,\l+\hat{i})|_{H},
\eea
where $H=Fun(W)[0]$  is preserved by the transfer matrices.

Let $f\longmapsto \int f(\l)$ be a linear functional on the space
${\cal F}$ of functions of $\l\in \Cb^n$, such that $\int
f(\l+\hat{i})=\int f(\l)$, $i=1,\cdots,n$, for all $f\in {\cal
F}$. Extend $\int$ to vector-valued functional by acting
component-wise. Then for each eigenfunction $\Psi(\l)$ of $T(u)$
with eigenvalue $t(u)$, the transfer matrices of $\Zb_n$ Belavin
model $T_{B}(u)$ act on the vector $\int S_m(z_1,\cdots,z_m;
\l)\Psi(\l)$
\bea
&&T_B(u) \int S_m(z_1,\cdots,z_m; \l)\Psi(\l)
=\sum_{i=1}^{n}\int S_m(z_1,\cdots,z_m;
\l+\hat{i})L^i_i(u,\l+\hat{i})\Psi(\l)\no\\
&&~~~~~~=\sum_{i=1}^{n}\int S_m(z_1,\cdots,z_m;
\l)L^i_i(u,\l)\Psi(\l-\hat{i})\no\\
&&~~~~~~=\int S_m(z_1,\cdots,z_m;\l)T(u)\Psi(\l)=t(u) \int
S_m(z_1,\cdots,z_m;\l)\Psi(\l). \eea Namely, the vector $\int
S_m(z_1,\cdots,z_m; \l)\Psi(\l)$ is an eigenvector of $T_B(u)$
with the same eigenvalue. Fortunately, such a linear functional on
${\cal F}$ has been given in \cite{Bax73, Fel96} for XYZ spin
chain ($\Zb_2$-Belavin model) and its  higher spin generalizations
\cite{Tak92}, for the generic $\Zb_n$ Belavin model \cite{Hou89}.
And our results recover those of  \cite{Fel96} for $n=2$ case and
coincide with those of \cite{Hou89} for $\Zb_n$ Belavin model.

As in \cite{Tak92}, one can construct the higher spin
generalization of $\Zb_n$ Belavin model by putting at each  site a
local $L$-operator
\bea
L_k(u;l_k)= \sum_{{\bf\a}\in
Z_n^2}\frac{\s_{{\bf \a}}(u-z_k+\frac{w}{n})}{\s_{{\bf
\a}}(\frac{w}{n})}I_{{\bf \a}}\otimes
  \rho_{l_k}\lt\{{\cal
S}_{{\bf \a}}\rt\},\quad l_k\in \Zb^+,\quad {\cal S}_{{\bf \a}}\in
SK, \eea where $\rho_{l_k}$ is the spin-$l_k$ representation of
$\Zb_n$ Sklyanin algebra $SK$ with dimension
$\frac{(n+l_k-1)!}{(n-1)!l_k!}$ \cite{Hou97}. The corresponding
transfer matrices of the model are given by \bea
T_B(u;\lt\{l_k\rt\})= tr_{0}\lt\{L_1(u;l_1)L_2(u;l_2)\cdots
L_m(u;l_m)\rt\}.\label{GeB} \eea Then, our construction in Sect.3
actually results in the eigenvalues formulation (\ref{t-function})
of the transfer matrices (\ref{GeB}), with the Bethe ansatz
equations (\ref{BAE}). In particular, our result recovers that of
\cite{Tak92} when $n=2$.

%%%%%%%%%%%%%%%%%%%%%%%%%%%%%%%%%%%%%%%%%%%%%%%%%%%%%%%%%%%%%%
%                                                            %
%The Elliptic $A_{n-1}$ Ruijsenaars-Schneider model           %
%                                                            %
%%%%%%%%%%%%%%%%%%%%%%%%%%%%%%%%%%%%%%%%%%%%%%%%%%%%%%%%%%%%%%
\section{The elliptic $A_{n-1}$ Ruijsenaars-Schneider model }
\label{RS} \setcounter{equation}{0}

If we take $\E$-module $W$ for the special case \footnote{Billey
used algebraic Bethe ansatz method over the $\E$-module
$W=V_{\L^{(1)}}(z_1)\otimes \cdots V_{\L^{(1)}}(z_1-(nl-1)w)$ to
obtain eigenvalues of elliptic N-body Ruijsenaars operator
\cite{Bil98}}
 \bea
W=V_{\L^{(l_1)}}(z_1),~~~\L^{(l_1)}=l_1\L_1=N\L_1=n\times l\L_1,
\eea the zero-weight space of this module is one-dimensional and
the associated {\it transfer matrices} can be  given by
\cite{Fel961} \bea T(u)=\frac{\s\left(\vTs
u-z_1+lw\right)}{\s(u-z_1+nlw)}M.\label{Ham} \eea The operator $M$
is independent of $u$ and is given by \bea M=\sum_{i=1}^n\lt\{
\prod_{j\ne i}\frac{\s(\l_{ij}w+lw)}{\s(\l_{ij}w)}\G_i\rt\}. \eea
Here $\lt\{\G_i\rt\}$ are elementary difference operators:
$\G_i\,f(\l)=f(\l-\hat{i})$. In fact, for a special choice of the
parameters \cite{Hou03}, this difference operator $M$ is  the
Hamiltonian of elliptic $A_{n-1}$ type Ruijsenaars-Schneider model
\cite{Rui86}
  with the special {\it coupling
constant\/} $\g=lw$, up to conjugation by a function \cite{Has97,
Hou00}. Now, we consider the spectrum of $M$. The results of Sect.
3, enable us to obtain the spectrum of the Hamiltonian of the
elliptic $A_{n-1}$ Ruijsenaars-Schneider model as well as the
eigenfunctions, in terms of  the associated {transfer matrices}
(\ref{Ham}). Since $M$ is independent of $u$,  we can evaluate the
eigenvalue of $T(u)$ at a special value of $u$, $u=z_1$. Then the
expression of the eigenvalue $t(u;\{v_k\})$ simplifies
drastically, for the second term in the right hand side of
(\ref{t-function}) (the one depending on the eigenvalue of the
reduced transfer matrices $t^{(1)}(u;\{v_k^{(1)}\})$) vanishes
because $u-p^{(0)}_1=0$ in the case (\ref{C}). Finally, we  obtain
the eigenvalues of $M$: \bea e^{(1-n)\bar{\a}
w}\frac{\s(nlw)}{\s(lw)}\lt\{\prod_{k=1}^{(n-1)\times l} \frac
{\s(v_k-z_1+w)} {\s(v_k-z_1)} \rt\}, \eea where $\{v_k\}$ and
$\{\{v^{(i)}_k\}\}$ are determined by the nested Bethe ansatz
equations \bea &&e^{(i-n)\bar{\a}^{(i)} w} \lt\{ \prod_{k=1,k\ne
s}^{N_{i+1}}\frac{\s(v^{(i)}_k-v^{(i)}_s+w)}{\s(v^{(i)}_k-v^{(i)}_s-w)}\rt\}=
\lt\{
\prod_{k=1}^{N_i}\frac{\s(v^{(i)}_s-p^{(i)}_k)}{\s(v^{(i)}_s-q^{(i)}_k)}\rt\}
t^{(i+1)}(v^{(i)}_s;\{v^{(i+1)}_k\}),\no\\
&&~~~~~~~~i=0,1,\cdots, n-2.\label{BAE2} \eea The functions
$t^{(i)}(u;\{v_k^{(i)}\})$ appearing in (\ref{BAE2})  are defined
by the following recurrence relations \bea
&&t^{(i)}(u;\{v^{(i)}_k\})= e^{(i+1-n)\bar{\a}^{(i)} w}
\lt\{\prod_{k=1}^{N_{i+1}}
\frac{\s(v^{(i)}_k-u+w)}{\s(v^{(i)}_k-u)}\rt\}\no\\
&&\qquad\qquad+e^{\bar{\a}^{(i)} w}
\lt\{\prod_{k=1}^{N_{i+1}}\frac{\s(u-v^{(i)}_k+w)}{\s(u-v^{(i)}_k)}\rt\}
\lt\{\prod_{k=1}^{N_i}\frac{\s(u-p^{(i)}_k)}{\s(u-q^{(i)}_k)}\rt\}
t^{(i+1)}(u;\{v^{(i+1)}_k\}),\no\\[2pt]
&&~~~~~~~~~~~~~~~~~~~~~~~~~~~~~~~~i=1,\cdots,n-2,\\
&&t^{(n-1)}(u)=1, \eea where $\bar{\a}^{(i)}$, $i=0,1,\cdots, n-2$
are given by (\ref{Relation}), $\bar{\a}=\bar{\a}^{(0)}$, $N_0=1$
and \bea
&&p^{(0)}_1=z_1,~~q^{(0)}_1=z_1-nlw,\label{C}\\
&&p^{(i)}_k=v^{(i-1)}_k,~~q^{(i)}_k=v^{(i-1)}_k-w,~~i=1,2,\cdots,
n-2,~~~k=1,2,\cdots,N_{i}.
\eea

%%%%%%%%%%%%%%%%%%%%%%%%%%%%%%%%%%%%%%%%%%%%%%%%%%%%%%%%%%%%%%
%                                                            %
%      Conclusions                                           %
%                                                            %
%%%%%%%%%%%%%%%%%%%%%%%%%%%%%%%%%%%%%%%%%%%%%%%%%%%%%%%%%%%%%%
\section{Conclusions}
\label{Con} \setcounter{equation}{0}

We have studied some modules over the elliptic quantum group
associated with $A_{n-1}$ type root systems. There are elliptic
deformation of symmetric tensor product of the fundamental vector
representation of $A_{n-1}$ (we have called it  {\it higher spin
representation}). Using the nested Bethe ansatz method, we exactly
diagonalize the commuting transfer matrices associated with the
tensor product of the $\E$-modules with generic evaluation points.
The eigenvalues of the transfer matrices and associated Bethe
ansatz equation are given by (\ref{t-function}) and (\ref{BAE}).
For the special case of $n=2$, our result recovers that of Felder
et al \cite{Fel96}.

We also take the applications of our construction to some
integrable models. For $l_1=l_2=\cdots=1$ case, our result gives
the spectrum problem of $\Zb_n$ Belavin model with periodic
boundary condition, which coincides with the result \cite{Hou89}.
For the special case $W=V_{\L^{(nl)}}(z_1)$, or equivalently
$l_1=\cdots=l_{nl}=1$ and $z_i=z_1-(i-1)w$, our result gives the
spectrum problem of the elliptic $A_{n-1}$ type Ruijsenaars
operators, which is ``relativistic" generalization \cite{Bra97} of
the Calogero-Moser type integrable differential operators, with
the special {\it coupling constant} $\g=lw$, which coincides with
the result \cite{Bil98}. Moreover, for the generic case, our
result gives eigenvalues of the {\it transfer matrices\/} given by
(\ref{GeB}) associated with higher spin generalization of $\Zb_n$
Belavin model with periodic boundary condition, and it further
enables us to construct the eigenvalues of all types of
Ruijsenaars operators associated with the $A_{n-1}$ root system in
the Bethe ansatz form \cite {Hou03}.

%%%%%%%%%%%%%%%%%%%%%%%%%%%%%%%%%%%%%%%%%%%%%%%%%%%%%%%%%%%%%%%
%                                                             %
%  Acknowledgments                                            %
%                                                             %
%%%%%%%%%%%%%%%%%%%%%%%%%%%%%%%%%%%%%%%%%%%%%%%%%%%%%%%%%%%%%%%
\section*{Acknowledgements}
We thank professors A. Belavin, S. Odake and K.J. Shi for their
useful discussion. This work is supported in part by Grant-in-Aid
for Scientific Research from the Ministry of Education, Culture,
Sports, Science and Technology, No.12640261. W.-L. Yang is
supported by the Japan Society for the Promotion of Science.

%%%%%%%%%%%%%%%%%%%%%%%%%%%%%%%%%%%%%%%%%%%%%%%%%%%%%%%%%%%%%%
%                                                            %
%     Appendix                                               %
%                                                            %
%%%%%%%%%%%%%%%%%%%%%%%%%%%%%%%%%%%%%%%%%%%%%%%%%%%%%%%%%%%%%%
\section*{Appendix A: The commutation relations}
\setcounter{equation}{0}
\renewcommand{\theequation}{A.\arabic{equation}}
The starting point for deriving the commutation relations among
$\A(u),~\D^j_i(u)$ and $\B_i(u)$ $(i,j=2,\cdots,n)$ is the
exchange relations (\ref{Exchange1}) and (\ref{Exchange2}). We
rewrite  (\ref{Exchange1}) in the component form
\bea
\sum_{i',j'=1}^{n}R^{a~b}_{i'j'}(u-v,\l-h) \tilde{L}^{i'}_c(u)
\tilde{L}^{j'}_d(v)= \sum_{i',j'=1}^{n}\tilde{L}^{b}_{j'}(v)
\tilde{L}^{a}_{i'}(u) R^{i'j'}_{c~d}(u-v,\l).\label{A1}
\eea
For $a=b=d=1,~c=i\ne 1$, we obtain
\bea \A(v)\B_i(u)&=&
\frac{1}{R^{i1}_{i1}(u-v,\l-\hat{1}-\hat{i})} \B_i(u)\A(v)-
\frac{R^{1i}_{i1}(u-v,\l-\hat{1}-\hat{i})}
{R^{i1}_{i1}(u-v,\l-\hat{1}-\hat{i})} \B_i(v)\A(u)\no\\
&=&
\frac{1}{R^{i1}_{i1}(u-v,\l)} \B_i(u)\A(v)-
\frac{R^{1i}_{i1}(u-v,\l)} {R^{i1}_{i1}(u-v,\l)} \B_i(v)\A(u),
\eea
by noting the definitions
(\ref{Elements1}) and (\ref{Elements2}).
The commutation relation (\ref{Rev1}) is a simple consequence
of the above equation.

Similarly, for $a=j\ne 1,~b=1,~c=i\ne 1,~d=l\ne 1$, we obtain \bea
\D^j_i(u)\B_l(v)=\sum_{\a,\b=2}^n\lt\{
\frac{R^{\a\b}_{i~l}(u-v,\l)}{R^{j1}_{j1}(u-v,\l-h)}
\B_{\b}(v)\D^j_{\a}(v)\rt\}- \frac{R^{j1}_{1j}(u-v,\l-h)}
{R^{j1}_{j1}(u-v,\l-h)} \B_i(u)\D^j_l(v), \eea from which follows
the commutation relation (\ref{Rev2}). For $a=b=1,~c=i\ne
1,~d=j\ne 1$, we obtain \bea
&&\B_i(u)\B_j(v)=\sum_{\a,\b=2}^{n}\B_{\b}(v)
\B_{\a}(u)R^{\a\b}_{ij}(u-v,\l)
=\sum_{\a,\b=2}^{n}R^{\a\b}_{ij}(u-v,\l-\hat{\a}-\hat{\b})
\B_{\b}(v)\B_{\a}(u)\no\\
&&~~~~~~~~~~~~~~=\sum_{\a,\b=2}^{n}R^{\a\b}_{ij}(u-v,\l)
\B_{\b}(v)\B_{\a}(u),
  \eea
  which leads to the commutation
relation (\ref{Rev3}). The other commutation relations
(\ref{Rev4})-(\ref{Rev5}) are derived from  (\ref{Exchange2}).

%%%%%%%%%%%%%%%%%%%%%%%%%%%%%%%%%%%%%%%%%%%%%%%%%%%%%%%%%%%%%%%
%                                                             %
%  References                                                 %
%                                                             %
%%%%%%%%%%%%%%%%%%%%%%%%%%%%%%%%%%%%%%%%%%%%%%%%%%%%%%%%%%%%%%%

\end{document}